\documentclass[smallcondensed]{svjour3}
\usepackage[utf8]{inputenc}
\usepackage{amsmath}
\usepackage{amsfonts}
\usepackage{amssymb}
\usepackage{graphicx}
\usepackage{dcolumn}
\usepackage[numbers]{natbib}
\begin{document}

% 	ETH Zurich\\
% 	Institute for Particle Physics\\
% 	CH-8093 Z\"urich\\
% 	Switzerland\\	

\title{Observation of positronium annihilation in the 2S state: towards a new measurement of the 1S-2S transition frequency}% Force line breaks with \\

\author{D.~A.~Cooke \and P.~Crivelli \and J.~Alnis \and A.~Antognini \and B.~Brown \and  S.~Friedreich \and A.~Gabard \and  T.~W.~Haensch \and K.~Kirch \and A.~Rubbia \and V.~Vrankovic}

\institute{
	D.~A.~Cooke
	\and
	P.~Crivelli 
	\and
	A.~Antognini
	\and
	S.~Friedreich
	\and
	A.~Rubbia
 \at
	ETH Zurich, Institute for Particle Physics, CH-8093 Z\"urich, Switzerland
	\and
	J.~Alnis \at
	Max-Planck-Institute of Quantum Optics, D-85741 Garching, Germany; \emph{present address:} University of Latvia, Riga, Latvia
	\and
	B.~Brown \at
	Marquette University, Milwaukee, WI 53233, USA
	\and
	T.~W.~Haensch \at
	Max-Planck-Institute of Quantum Optics, D-85741 Garching, Germany
	\and
	K.~Kirch \at
	ETH Zurich, Institute for Particle Physics, CH-8093 Z\"urich, Switzerland/Paul Scherrer Institute, CH-5232 Villigen, Switzerland
	\and
	A.~Gabard 
	\and
	V.~Vrankovic\at
	Paul Scherrer Institute, CH-5232 Villigen, Switzerland\\
	\email{paolo.crivelli@cern.ch}
}
	
\date{Received: date / Accepted: date}

\maketitle

\begin{abstract}
We report the first observation of the annihilation of positronium from the 2S state. Positronium (Ps) is excited with a two-photon transition from the 1S to the 2S state where its lifetime is increased by a factor of eight compared to the ground state due to the decrease in the overlap of the positron electron wave-function. The yield of delayed annihilation photons detected as a function of laser frequency is used as a new method of detecting laser-excited Ps in the 2S state. This can be considered the first step towards a new high precision measurement of the 1S--2S Ps line.
% \keywords{keyword1 \and keyword2}
% \PACS{pacs1 \and pacs2}
\end{abstract}

\section{Introduction}
Positronium (Ps) is the bound state of the positron and electron. Free from the finite-size effects that become increasingly significant in high-precision spectroscopy of H \citep{Beyer2013}, Ps is a purely leptonic atom for which weak interaction effects are negligible at the current level of experimental accuracy, and thus provides an excellent testing ground for bound-state QED \citep{Karshenboim2005}. The energy levels of Ps can be calculated perturbatively as an expansion of the fine structure constant $\alpha$ with very high precision only limited by the complexity of the calculations and the knowledge of the fundamental constants. This is not the case, for instance, for the hydrogen atom where the main source of the uncertainty in the predictions (involving S-states) are limited by the finite size of the proton. In comparison to other hydrogen-like atoms, however, annihilation  must be taken into account and recoil effects are not suppressed by the $m/M$ factor. The same theoretical tools could also be applied in the field of quantum chromodynamics (QCD) for the calculation of the energy levels of quarkonium (the bound state of 2 quarks) \citep{Czarnecki2001}. High-precision spectroscopy of Ps also allows the positron/electron mass ratio to be determined. This, together with other precision experiments with Ps such as searches for `invisible' decays, which may indicate a hidden mirror particle sector (e.g. \citep{Crivelli2010a} and references therein), could shed some light on the apparent matter/antimatter asymmetry in the Universe, and perhaps on the nature of dark matter.\par

Positronium has two ground states: the triplet ($1^3S_1$, ortho-positronium) and the singlet ($1^1S_0$, para-positronium) spin states.  Due to the odd-parity under C-transformation, ortho-positronium decays predominantly into three photons with a lifetime in vacuum of $\tau_{ops}=142.05$ ns \citep{Adkins2002,Kataoka2009}. Para-positronium decays predominantly into two photons with a lifetime in vacuum of $\tau_{pps }=125$ ps \citep{Czarnicki1999,AlRamadhan1994}. In the following, we will solely consider ortho-positronium (Ps) that, because of its longer lifetime (due to the phase-space and additional $\alpha$ suppression factors), allows us to perform the proposed studies.\par

This paper reports on the first detection of laser-excited 2S state Ps by using the difference in lifetime between the ground and excited states. This can be considered as the first step towards a new high precision measurement of the 1S--2S Ps line. This quantity has previously been measured using a pulsed laser \citep{Chu1984} and then by using a continuous wave (CW) laser \citep{Fee1993}. These measurements used an additional laser to photoionize the resulting excited-state Ps in order to detect the 2S state. Reaching a precision of 2.6 ppb, they provided the current limit on the electron/positron mass ratio and tested bound-state QED calculations up to $O(\alpha^4)$, as calculated by \citep{Fell1992}. The most recent calculations are up to $O(\alpha^7)$ \citep{Melnikov1999,Pachucki1999}, so an updated experimental determination is timely.

\section{Experimental method}

\begin{figure}
\centering
\includegraphics[width=0.8\textwidth]{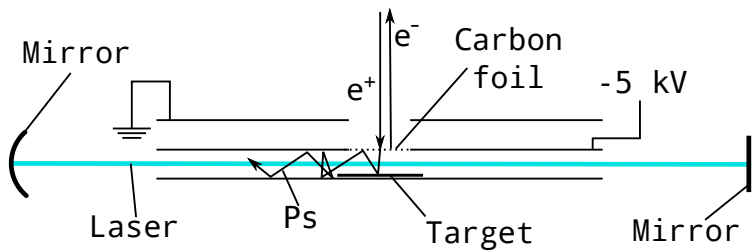}
\caption{Target geometry.}
\label{fig:target}
\end{figure}

\begin{figure}
\centering
\includegraphics[width=0.6\textwidth]{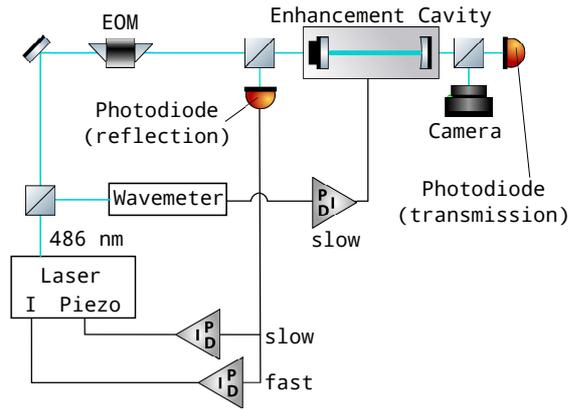}
\caption{Laser frequency stabilization scheme.}
\label{fig:lock}
\end{figure}

The measurement utilizes the ETHZ positron beam, which has been described previously \citep{Alberola2006}, with modifications. Briefly, the positrons are derived from a $^{22}$Na source coupled to a solid rare gas moderator system. Positrons with energy of 200 eV are magnetically transported around a 90$^\circ$ bend (to reduce contamination from fast particles and remove the direct line-of-sight between the source and the detectors) to the target region. The target is formed by a 4 $\mu g/m^2$ (about 20 nm thick) carbon foil with a porous silica thin-film positron--positronium converter 3 mm behind it. The whole assembly is biased at $-5$ kV, so positrons are implanted with slightly less than 5 keV kinetic energy (owing to energy loss in the window). Secondary electrons from the C-foil are accelerated by this voltage back along the beamline where they are detected on a multichannel plate, allowing the implanted positrons to be time-tagged. A significant fraction of the positrons implanted in the porous film are re-emitted into vacuum, bound as Ps, with near thermal energies \citep{Crivelli2010,Cassidy2010}. The target is fixed near the centre of a high-finesse ($F \sim 2.25\times10^5$) Fabry-Perot laser enhancement cavity such that the laser passes between the window and silica converter (as shown in Figure \ref{fig:target}).\par

The laser itself is a frequency-doubled diode laser system from Toptica producing up to $\sim 750$ mW of 486 nm light. In order for spectroscopy to be feasible the laser power must be enhanced to $\sim 1000$ W to achieve intensities of the order of 1 MW/cm$^2$. To reach this value, approximately 500 mW of 486 nm light is injected into the high-finesse cavity with an in-coupling efficiency of $\sim 15\%$. The enhancement cavity comprises two supermirrors (curvature radius 2 m and infinite for output and input mirrors respectively, transmission and absorption both 7 ppm) separated by 30 cm, which leads to an enhancement factor of $\sim 1\times10^4$. Since, the beam waist at the centre of the cavity is $\sigma=370 \mu$m, the separation between the laser beam axis and the target is therefore approximately 5 times the beam waist, so that the enhancement factor of the cavity is unaffected by the loss of power due to the laser touching the e$^+$/Ps converter. Although the advertised damage threshold of these mirror should have allowed a circulating power of 1500 W, degradation of cavity performance was observed already at 700 W. The circulating power was therefore deliberately limited to 500 W to avoid further damage to the mirrors.\par

To compensate for the power limitation of the mirrors, a new kind of target geometry was designed and implemented  (see Fig. \ref{fig:target}). The intention is to confine Ps spatially so that the Ps atoms could be redirected from the walls into the laser beam. In fact as Ps is very light, even at room temperature it is very fast ($v \approx 10^4$ m s$^{-1}$), so the interaction time with the laser is correspondingly short (on the order of 10 ns). Spatial confinement results in an estimated enhancement of a factor of five. To assess the effect of the collisions of Ps in its ground state with the walls of the tube, we measured the lifetime and intensity of the Ps emitted into vacuum for three target--window separations (3 mm, 5 mm and 9 mm) and observed no variation in either property with distance.  Destructive processes which may occur at the walls are limited to annihilation, as there is insufficient kinetic energy for Ps break-up. The annihilation of the positron in the Ps atom with an electron which is not its bound partner (so called pick-off effect) is proportional to the local electron density, thus, the cross-section for this process will be  proportional to the time spent in the locality of the wall, or inversely proportional to the velocity. Our measurements suggest that the process is negligible at these energies.\par

For this measurement, the frequency is stabilized using the locking scheme shown in Figure \ref{fig:lock}, that is, with the laser locked to the enhancement cavity. The medium-term stability of 10 MHz precision was achieved by controlling the length of the enhancement cavity via a HighFinesse WS7 wavemeter. Long term stability can be monitored by Te$_2$ saturated spectroscopy, using a line close to the Ps line \citep{McIntyre1986}.\par

Detection of 2S Ps is achieved by exploiting the greatly increased lifetime of ortho-Ps in the 2S state compared to the ground state (1.1 $\mu$s vs 142 ns). Two large solid angle BGO scintillator arrays (comprising four $20\times6$ cm crystals each), placed above and below the cavity, are used to detect the annihilation quanta from positronium formed in the converter. A signal event is defined as the detection of an energy deposition between 100 keV--600 keV within 10 ns in two BGO crystals in a time window of 2--4 $\mu$s after the tagging time. This time window is chosen to maximize the signal to background ratio, since the decay of ground state positronium after 2 $\mu$s are suppressed to a level below 1 ppm ($e^{-2\mu s/0.142\mu s}$). The other sources of background are events that are not correlated with the tagged positron. A beam-related background arises in the case that a positron which is not the tagged one produces annihilation photons detected within the signal window. This effect is greatly reduced by biasing off the positron beam with 220 V on an electrode (the beam energy is 200 eV), following the detection of secondary electrons. Accidental background not related to the beam is generated by cosmic rays, environmental radioactivity and photomultiplier noise and is reduced by requiring that the energy and time cuts described previously. To reduce the effect of internal radioactivity, the BGO crystals are shielded from each other by 25 mm of lead. These measures reduce the accidental rate to a level of $<10$ events per $10^7$ triggers. For comparison, simulations of expected signal rate yield a value of 28 events per $10^7$ triggers, assuming a 10\% positron-to-positronium conversion efficiency and for 500 W of laser power on resonance.

\section{Resonance curve}
\begin{figure}
\centering
\includegraphics[width=0.6\textwidth]{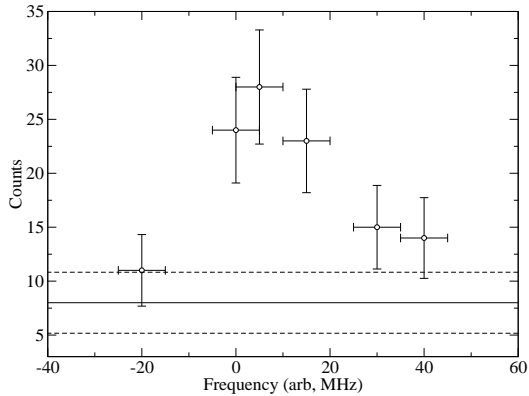}
\caption{Number of observed events in $10^7$ triggers vs. wavemeter frequency.}
\label{fig:curve}
\end{figure}

Figure \ref{fig:curve} shows the resonance curve for laser excitation of Ps 1S $\to$ 2S, as measured using the lifetime method described above. The peak of the curve is 28 events on a background level of 8, in reasonable agreement with the expected signal rate for the laser power achieved during the measurement---$\sim 500$ W---using a positron-to-Ps conversion efficiency of $\sim 10$\%. The position of the maximum of the resonance is arbitrary as reported by the WS7 wavemeter: this is not an absolute frequency measurement. Time-of-flight broadening of approximately 30 MHz is evident. The measurements were recorded in a time period of approximately four hours, and a calibration check against a Te$_2$ reference was performed before and after, to ensure that the laser frequency had not drifted significantly over the measurement.\par

The broad agreement between the peak signal rate and the prediction from simulation suggests that 2S Ps also survives collisions with the walls of the tube. The simulation was performed assuming 100 \% (diffuse) reflection from the surfaces. In addition to pick-off annihilation, 2S Ps may also be expected to undergo collisional de-excitation, mainly by mixing with the 2P state (which subsequently decays rapidly by emission of a photon). That simulation and experiment agree without the inclusion of this effect implies that the energy scale where collisional de-excitation becomes significant is below that typical of room temperature.\par

The observation that thermal energy Ps appears to survive such collisions largely unscathed suggests this technique for Ps confinement could be used in other relevant experiments, for example in Ps production for the GBar collaboration \citep{Perez2012}, or for studies of Ps--surface interactions. It is also in broad agreement with the calculations of \citep{Starrett2008} for excited-state Ps--atom interaction cross-sections, where the lowest energy cross-section for collisions with atomic Ar are $<1\times10^{-16}$ cm$^2$ (below 0.5 eV). That slow Ps can survive multiple collisions with silica surfaces is perhaps not surprising. In order to escape from the porous silica target, the Ps must diffuse through the pore network, undergoing numerous collisions at roughly the same energy, so the very fact that Ps can be emitted into vacuum from these materials suggest a high collision survival rate.\par

At 5 keV implantation energy, the implantation depth is such that positronium has thermalized within the material prior to emission. A conductive window is required to shield the spectroscopy volume from electric field penetration, for this measurement, 20 nm carbon foils were used. These foils are extremely delicate, however, and it is planned that the window material be replaced by gold-coated silicon nitride for greater structural integrity. The energy loss spectrum of positrons transmitted through an uncoated $3\times3$ mm, 30 nm thickness silicon nitride window was measured to ensure that the positrons retain enough kinetic energy to be sufficiently implanted. Above approximately 2 keV impact energy the transmitted fraction reaches a saturation value, with the energy loss peaked at 100--200 eV.

\section{Outlook}
This measurement constitutes a preliminary step towards a new determination of the Ps 1S-2S transition frequency, using the long lifetime of 2S Ps to detect excited Ps atoms. The experimental set-up is currently being reconfigured to allow to reach the aimed precision of $5\times 10^{-10}$ in order to check the recent QED calculations \citep{Melnikov1999,Pachucki1999}. This involves moving from using a time-tagged, continuous positron beam, to using the pulsed (and possibly bunched, i.e. with the pulses time-compressed) output from a two-stage buffer gas trap \citep{Surko2000}. Pulsed-beam operation will effectively eliminate the background arising from untagged positrons and accidentals, thus improving the S/N ratio of our method by two orders of magnitude. With sufficient events, the Ps 2S lifetime could be determined, which would serve as a measure of the stray electric field in the target tube (which should be $< 10$ Vcm$^{-1}$). Furthermore, the lower background will give us the possibility to use simultaneously a second method (already implemented in our set-up but not used due to the high beam related background) for the detection of the 2S states based on three photons resonant ionization of the Ps atoms in the same excitation laser. This technique can potentially yield additional information about the kinetic energy of the Ps atom, allowing for correction of the resulting lineshape for the 2nd-order Doppler shift.\par

Significant improvements in the precision of the 1S$\to$2S spectroscopy measurement using Ps emitted from silica thin-films used as is are unlikely as even at thermal energies the velocity of Ps is large. Possibilities for lowering the energy further are laser cooling and Stark deceleration of Rydberg Ps atoms, which are variously discussed in \citep{Liang1988, Hogan2013} and with specific reference to spectroscopy in (\citep{Crivelli2014,CrivelliGranit2014}. Stark deceleration of Rydberg Ps, followed by de-excitation back to the ground state could, in principle, allow the production of 1S Ps at rest and therefore allow for a much higher accuracy that might result in a independent determination of the Rydberg constant free from finite size effects.\par

\begin{acknowledgements}
This work was supported by the Swiss National Science Foundation under the Ambizione grant PZ00P2\_132059 and ETH Zurich Research Grant ETH-47-12-1. We would like to thank the IPP institute and ETH Zurich for their support, A. Gendotti, M. Wymann and the IPP workshop for their help with construction, S. Haas for assistance with data acquisition, and the PSI Magnet group for construction and characterization of the cavity coils.
\end{acknowledgements}

\bibliographystyle{spphys}
\bibliography{bibliography.bib}

\end{document}